\documentclass[conference]{IEEEtran}
\IEEEoverridecommandlockouts
\usepackage{cite}
\usepackage{amsmath,amssymb,amsfonts}
\usepackage{algorithm}
\usepackage{algpseudocode}
\usepackage{graphicx}
\usepackage{textcomp}
\usepackage{xcolor}
\usepackage{svg}
\usepackage{import}
\usepackage{tikz}
\usepackage{pgfplots}
\usepackage{comment}
\usepackage{environ}
\def\BibTeX{{\rm B\kern-.05em{\sc i\kern-.025em b}\kern-.08em
    T\kern-.1667em\lower.7ex\hbox{E}\kern-.125emX}}

\DeclareUnicodeCharacter{2212}{\textendash}

\renewcommand{\figurename}{Fig.}
\newcommand{\figref}[1]{\figurename~\ref{#1}}

\makeatletter
\newcommand*{\centerfloat}{%
  \parindent \z@
  \leftskip \z@ \@plus 1fil \@minus \textwidth
  \rightskip\leftskip
  \parfillskip \z@skip}
\makeatother

\begin{document}

\title{An Algorithm for Exact Numerical Age-of-Information Evaluation in Multi-Agent Systems
\thanks{
RIchard Schöffauer is supported by the DFG under priority programm SPP 1914. Gerhard Wunder is supported by both the DFG (SPP 1914) and the BMBF (6G-RIC).}
}

\author{\IEEEauthorblockN{RIchard Schöffauer}
\IEEEauthorblockA{\textit{Dept. of Mathematics and Computer Science} \\
\textit{Freie Universität Berlin}\\
Berlin, Germany \\
richard.schoeffauer@fu-berlin.de}
\and
\IEEEauthorblockN{Gerhard Wunder}
\IEEEauthorblockA{\textit{Dept. of Mathematics and Computer Science} \\
\textit{Freie Universität Berlin}\\
Berlin, Germany \\
gerhard.wunder@fu-berlin.de}
}

\maketitle

\begin{abstract}
We present an algorithm for the numerical evaluation of the state-space distribution of an Age-of-Information network. Given enough computational resources, the evaluation can be performed to an arbitrary high precision. An Age-of-Information network is described by a vector of natural numbers, that track how outdated status information from various agents is. Our algorithm yields the means to determine any moment of the corresponding stochastic process. This can be extremely valuable for cases in which the network consists of controllers that communicate with one another, as it potentially allows for less conservative control behavior. It also enables the comparison of different policies regarding their performance (minimizing the average Age-of-Information) to a much more accurate degree than was possible before. This is illustrated using the conventional MaxWeight policy and the optimal policy. We also validate and compare the algorithm with Monte-Carlo-Simulations.
\end{abstract}


\section{Introduction}

In recent years, the so-called Age-of-Information (AoI) metric has gained considerable attention in lieu of the conventional Communication-Delay (ComDelay) metric.
While ComDelay captures the elapsed time between transmission and reception, AoI measures the age of information at the receiver. To appreciate the difference, imagine a communication line with a 1 second ComDelay and assume that a packet of information is send over that line only every full hour. The ComDelay is not influenced by this usage and remains 1 second; however the AoI at the receiver is about 30 minutes on average. It starts from 1 second every full hour and peaks at 1 hour and 1 second just before the next reception.

In contrast to ComDelay, AoI is especially well suited towards the needs of the information-receiving agents in a network. And in cases where the receiving agent is a controller awaiting new status-updates on its input variables (i.e. when the network closes a control loop), AoI is the much more significant indicator for the overall performance. While AoI grows linearly in time, it is reset abruptly to one (or zero, depending on the context), once a new status-update is received. This non-linear behavior is the reason why analytical results on the stochastic process describing the AoI are rare and usually of limited quality.

In search for an optimal scheduling policy, i.e. a policy that minimizes the average AoI, \cite{Kadota2016} shows that 
a greedy policy is optimal, but only in the case of symmetric networks. For the general case, \cite{Hsu2017} derives requirements on the optimal policy by modeling the network as a Markov-Decision-Process. \cite{Hsu2020} shows, that the same approach allows to employ restless bandit methods for leveraging a well performing policy.
In \cite{Kadota2019, Kadota2018, Kadota2018a}, the authors develop a lower bound on the average AoI and manage to link the performance of multiple low-complexity policies to that bound. However, the results are quite weak; e.g. while the MaxWeight policy seems to yield an average AoI very close to this lower bound in simulations, analytically it can only be guaranteed to yield less than double that bound.

AoI in networks with explicit multi-hop propagation is investigated in \cite{Arafa2017, Yates2018}, where the focus lies on specific topologies (e.g. line or ring networks). For a general topology, fundamental bounds on average and peak AoI are derived in \cite{Farazi2019}.

As mentioned, AoI is especially well suited if the network is part of a closed control loop. In \cite{Hahn2021}, the authors show how forecasting AoI values can improve the performance of model-predictive controllers with a common control goal. For cellular networked control systems, \cite{Ayan2019} compares the AoI metric with a Value-of-Information metric which considers the expected information content of an update. The most practical investigation is performed in \cite{Sinha2019}, where the authors even consider processing delay and physical execution times for industrial wireless sensor-actuator networks, in order to derive policies that minimize the average AoI.

Our \textbf{contribution} makes it possible to derive an exact numerical ratio between the performance of the optimal (average AoI minimizing) policy and the MaxWeight policy in the case of a two-agent network. In addition to that, a straightforward extension of the presented algorithm allows to yield such results for any policy and any number of agents (only limited by computational resources). Since the algorithm computes the exact probability distribution over the state-space, it enables system designers to derive the stochastical moments of the network process, facilitating a more accurate prediction of the network performance (and in cases where the network is part of a control loop a less conservative parametrization of the distributed controllers). In this paper, we present the algorithm including its application to the MaxWeight and the optimal policy.

\section{System  Model}

The underlying system model for this paper is inspired by the use case of distributed controllers, communicating in order to achieve a common control goal. Each controller operates autonomously and thus naturally generates status-updates of all relevant local quantities. Its individual control goal, however, is influenced by other quantities, only observable (because local) to other controllers in the network. Outdated information of those quantities directly diminishes the controller's performance. Hence, each controller is \textit{always} motivated to broadcast its freshly generated status to all other controllers.

In particular, we assume a network of $n \in \mathbb{N}$ agents who communicate over a wireless resource such that only one agent may broadcast at any given time. (Here and throughout the paper, $\mathbb{N}$ does \textit{not} contain $\{0\}$!). Time is slotted and in each slot, every agent generates an update of its own status that can potentially be broadcasted to all other agents. Every agent remembers the latest received status information of all other agents. Let the AoI of agent's $i$ last broadcasted status-update at all other agents be denoted by $a_t^{i} \in \mathbb{N}$. The collection $(a_t^1, a_t^2,\dots a_t^n) \in \mathbb{N}^n$ therefore fully describes the network state in time-step $t$. Given no updates, $a_t^{i}$ increases linearly with time but is \textit{reset} to one, once a status-update is received. In order for this to happen, in each time-step a policy determines which agent is to broadcast its current status to all other agents and thus which AoI component is to be reset. If agent $i$ is to broadcast in time-step $t$ then the control variable $v^i_t \in \{0,1\}$ is one, otherwise it is zero. The resource constraint requires that $\sum_i v_t^i = 1$. Whether such a transmission from agent $i$ to all other agents would succeed is determined by the stochastic variable $p_t^i \in \{0,1\}$, which is one in case of success and zero otherwise. The sequence $\{p_t^{i}\}$ describes a Bernoulli process with success probability $p^{i} \in [0,1] \subset \mathbb{R}$ (failure probability $\bar{p}^i = 1 - p^i$) and serves to model channel fading and other stochastic disturbances.
Put together this results in the following evolution of the AoI:
\begin{equation}
    \label{eq::evolution}
    a_{t+1}^{i} = 1 + a_t^{i} \left( 1 - v_t^i p_t^{i} \right)
\end{equation}

With with slight abuse of notation, we let $a \in \mathbb{N}^n$ denote a state of the network and $a^i$ ($i \in \{1,\dots n\}$) its components, independent of time. Since the system dynamics are time invariant, we need only consider stationary policies that map each state $a$ to a network agent $i$. Such a mapping, called a decision $d$ fully defines a policy:
\begin{equation}
    d: \mathbb{N}^n \to \{1, \dots n\}
\end{equation}
Furthermore, we will only consider causal policies, i.e. policies for which holds
\begin{equation}
    d(a) = i \quad \Longrightarrow \quad d(a+e^i) = i
\end{equation}
with $e^i$ being the $i$-th unit vector. I.e. given a state $a$ and the policy's decision $d(a) = i$, the policy's decision stays on agent $i$, if we were to increase the AoI only in the $i$-th component. Most reasonable policies (like MaxWeight) fulfill this property.


\section{Methodology for Evaluation of Exact State-Space Distribution}

\subsection{Notation \& Technique}

Let $f(a)$ be the probability, with which the network is in state $a$, given an arbitrary time-slot and no prior information. To evaluate $f(a)$ we utilize the special structure of the system evolution \eqref{eq::evolution}: Every state $a$, \textit{not} at the boundary of the state-space can only be reached from its diagonal predecessor ${a - \mathbf{1} = (a^1-1,\dots \ a^n-1)}$, and only if in this predecessor-state, the intended reset did fail (which happens with probability $\bar{p}^{d(a-\mathbf{1})}$). The sequence of consecutive predecessors of such a state form a diagonal line that ends at the state
\begin{equation}
\begin{gathered}
    a - \mathbf{1} \cdot \left( \min_i\{a^i\} + 1\right) 
    \\ = \left( a^1-\min_i\{a^i\}+1,
    \dots \ a^n-\min_i\{a^i\}+1 \right)
\end{gathered}
\end{equation}
at the boundary of the state-space (dark-green states in \figref{fig::MW_state_space}). We call this the "root-state" of $a$. Once the probabilities of all root-states are known, the remaining distribution over the state-space follows immediately by construction of said diagonals.

The policy determines how probability diminishes along a diagonal: The product of all failure probabilities $D(a)$, necessary to reach state $a$ from its root-state, can be expressed recursively by
\begin{equation}
    D(a) =
    \begin{cases}
        1 & \text{if } \min_i \{a^i\} = 1
        \\
        D(a - \mathbf{1}) \cdot \bar{p}^{d(a-\mathbf{1})} & \text{otherwise}
    \end{cases}
\end{equation}
Hence, we have the following connection between a state's probability and its root probability:
\begin{equation}
    \label{eq::diagonal_attenuation}
    f(a) = D(a) f(a - \mathbf{1} \cdot \left( \min_i\{a^i\} + 1\right))
\end{equation}
Finally, given any causal policy, a root-state, let's say $(1,a^2,\dots \ a^n)$, can be expressed by
\begin{equation}
    \label{eq::general_reset}
    f(1,a^2,\dots\ a^n) = p^1 \sum_{a^1 = b}^\infty f(a^1,a^2-1,\dots \ a^n-1)
\end{equation}
where $b$ is determined by the employed policy. In particular, we say that state $(b,a^2-1,\dots \ a^n-1)$ is the "first" state, from which $(1,a^2,a^3,\dots \ a^n)$ is reachable. A visual representation of \eqref{eq::general_reset} are the light-blue and light-green sets of states in \figref{fig::MW_state_space}.

\subsection{Evaluation of the State-Space Distribution under the MaxWeight Policy for $n=2$}

\begin{figure}
    \centering
    \footnotesize
    \includegraphics[width = 0.75\columnwidth]{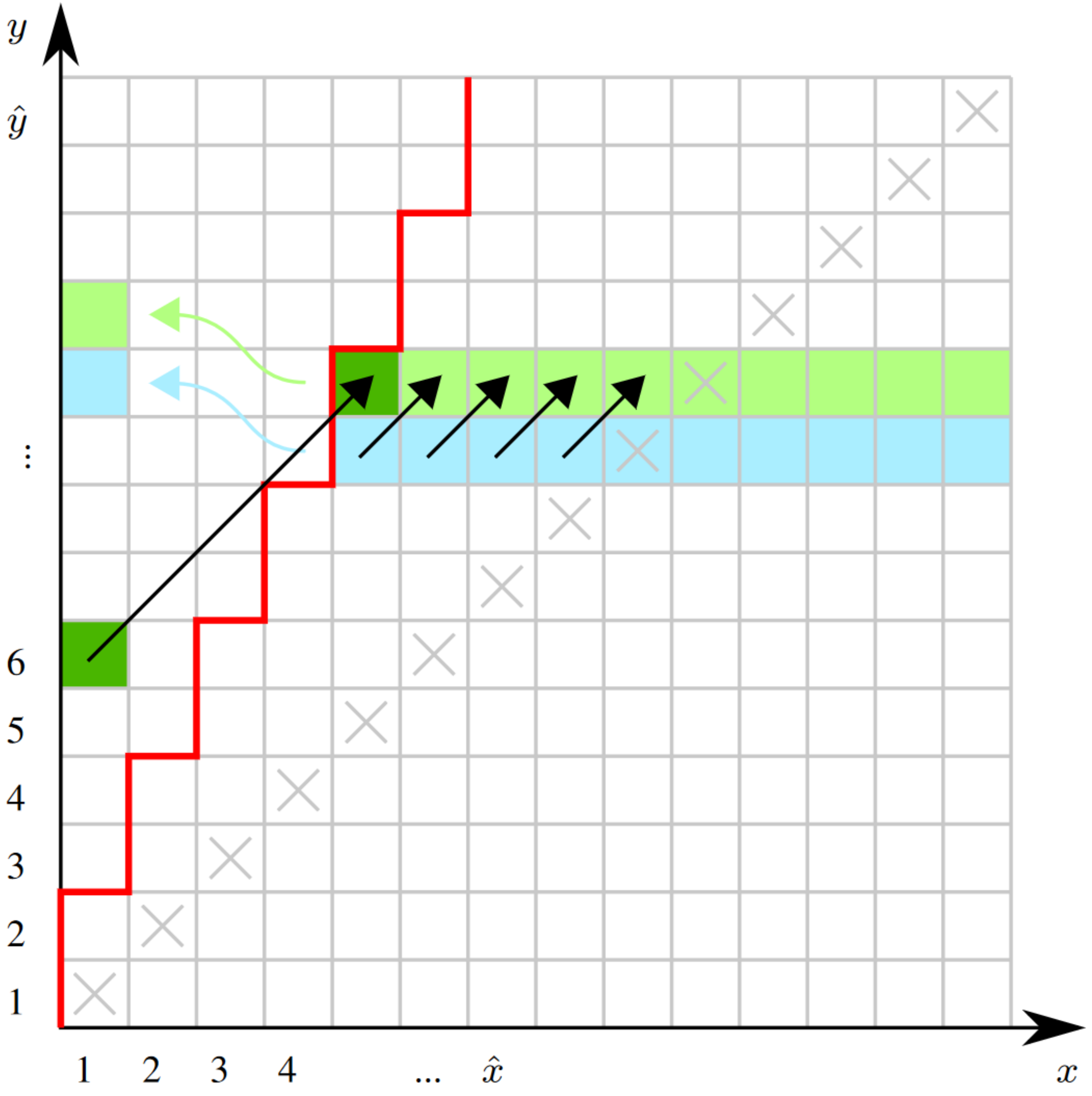}
    \caption{Root-state recursion under MW policy in 2-dim. state-space. States on the boundary (root-states) are only reachable from the highlighted sets of states (light-green / blue). States not on the boundary are only reachable from their diagonal predecessor.}
    \label{fig::MW_state_space}
\end{figure}


\newsavebox\proma
\savebox\proma{%
\begin{tikzpicture}
\begin{axis}[
    colormap/cool,
    view = {135}{45},
    width = 55mm,
    height = 55mm,
    font = \footnotesize,
    xticklabels={},
    yticklabels={},
    ztick={0,0.05},
    zticklabels={ $0$ , $0.05$ },
    ztick scale label code/.code={},
    z tick label style={rotate=90,anchor=south},
    zmin = {0},
    zmax = {0.06}
    ]
\addplot3 [surf, mesh/ordering=y varies] table {data_proma_p06_q02.txt};
\end{axis}%
\end{tikzpicture}
}
\newsavebox\incma
\savebox\incma{%
\begin{tikzpicture}
\begin{axis}[
    colormap/cool,
    view = {135}{45},
    width = 55mm,
    height = 55mm,
    font = \footnotesize,
    xticklabels={},
    yticklabels={},
    ztick={0,0.05},
    zticklabels={ $0$ , $0.05$ },
    ztick scale label code/.code={},
    zticklabel pos = right,
    z tick label style={rotate=90,anchor=north},
    zmin = {0},
    zmax = {0.06}
    ]
\addplot3 [surf, mesh/ordering=y varies] table {data_incma_p06_q02.txt};
\end{axis}%
\end{tikzpicture}
}
\newsavebox\difma
\savebox\difma{%
\begin{tikzpicture}
\begin{axis}[
    colormap/cool,
    view = {135}{45},
    width = 55mm,
    height = 55mm,
    ztick={0,0.0002},
    zticklabels={ $0$ , $0.0002$ },
    ztick scale label code/.code={},
    font = \footnotesize,
    xlabel style = {sloped},
    ylabel style = {sloped},
    xlabel={$x$-component of AoI},
    ylabel={$y$-component of AoI},
    zlabel={probability*},
    z tick label style={rotate=90,anchor=south},
    zmin = {0},
    zmax = {0.0004}
    ]
\addplot3 [surf, mesh/ordering=y varies] table {data_difma_p06_q02.txt};
\end{axis}%
\end{tikzpicture}
}

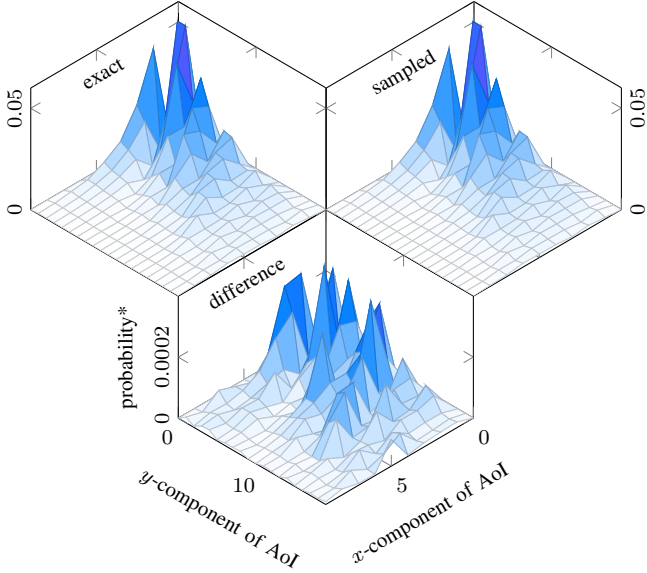
\begin{figure}
    \centering
    \footnotesize
    \begin{tikzpicture}[x=1mm,y=1mm] 
        \node at (0,-33) {\usebox\difma};
        \node at (-19.5,0) {\usebox\proma};
        \node at (22,0) {\usebox\incma};
        \node[rotate=30] at (-29,10) {exact};
        \node[rotate=30] at (11,10) {sampled};
        \node[rotate=30] at (-10,-17.8) {difference};
    \end{tikzpicture}
    \caption{Probability distribution over the state-space for the MW policy: Exact values following the described method (upper left); Sampled values through $10^6$ simulation steps (upper right); Absolute difference between the two (down mid).}
    \label{fig::mw_triple}
\end{figure}

Given these tools, we will now investigate the AoI under the well known MaxWeight (MW) policy, in the 2 dimensional case. The methodology is applicable to higher dimensional cases, albeit with slightly more linear algebra in order to determine the required variables. For succinct notation we will identify $a^{1}$ and $a^{2}$ with $x$ and $y$, and $p^{1}$ and $p^{2}$ with $p$ and $q$, respectively. Also, $b$ from \eqref{eq::general_reset} will be denoted by $x'(y)$ or $y'(x)$, referring to the $x$ or $y$ component of the first state, from which the root-state in question is reachable. We can specify these functions given the employed policy: MW separates the state-space linearly into two halves. Whenever $x p \geq y q$, the system will try to reset $x$ with probability $p$, otherwise $y$ with probability $q$.
Hence the first state from which $(1,y)$ is reachable is $(x'(y-1),y-1)$ with
\begin{equation}
    x'(y) = \left\lceil \frac{q}{p}y \right\rceil
\end{equation}
since $x'(y)p \geq yq$; and vice versa for root-states of the form $(x,1)$.

W.l.o.g. we will assume that $p > q$. Then, as visualized by \figref{fig::MW_state_space} (blue states), we can express the probability of a root-state on the $y$-axis as
\begin{equation}
\label{eq::twosums}
\begin{aligned}
    f(1,y+1) &= p \sum_{x = x'(y)}^\infty f(x,y)
        \\ &= p \sum_{x=x'(y)}^{y} f(x,y) + p \sum_{x=y+1}^\infty f(x,y)
        \\ &= p \sum_{x=x'(y)}^{y} f(1,y-x+1) D(x,y)
        \\ & \qquad + p \sum_{x=y+1}^\infty f(x-y+1,1) D(x,y)
\end{aligned}
\end{equation}
Crucially, no matter the choice of $y$, the last sum will always originate from the probability of all the root-states on the $x$-axis. In contrast, the first sum refers back to root-states on the $y$-axis that came before the state $(1,y)$, constituting a recursion.

To quickly yield this recursion, we compare \eqref{eq::twosums} for states $(1,y+1)$ and $(1,y)$:
The probability of reaching $(1,y+1)$ is equal to the probability of reaching $(1,y)$, attenuated by the probability that reaching $(1,y)$ did fail (diagonal shift from blue to green states in \figref{fig::MW_state_space}). However, from time to time, $(1,y+1)$ can also be reached from an additional state whose predecessor does not stem from the states that $(1,y)$ was reachable from (dark green state in \figref{fig::MW_state_space}). This "disturbance" in the recursion is caused by a change in $x'(y)$ and thus directly connected to the employed policy. The root of these additional states must be on the $y$-axis.

In particular, If $x'(y)$ is not $x'(y-1) + 1$, there must be an additional state from which $(1,y+1)$ is reachable, namely $(x'(y),y)$. The root of this state is $(1,y-x'(y)+1)$, and according to \eqref{eq::diagonal_attenuation} we have
\begin{equation}
\begin{aligned}
    f(x'(y),y) &= f(1,y-x'(y)+1) D(x'(y),y) 
    \\& = f(1,y-x'(y)+1) \bar{q}^{x'(y)-1}
\end{aligned}
\end{equation}
This leads to the recursive formula
\begin{equation}
    \label{eq::recursion}
\begin{gathered}
    f(1,y+1) = \bar{p} f(1,y) + p \cdot \dots
\\[2ex]
    \dots \begin{cases}
    f(1,y-x'(y)+1) \bar{q}^{x'(y)-1}
    & \text{if } x'(y) = x'(y-1)
    \\
    0 & \text{otherwise}
    \end{cases}
    \end{gathered}
\end{equation}

With that we are just missing an initial value to start the recursion. Naturally we start with
\begin{equation}
    f_A(1,2) = p \sum_{x = x'(1)}^\infty f_A(x,1) = 1
\end{equation}
since $f(1,1) = f(2,2) = \dots = 0$ ($x$ and $y$ cannot be reset at the same time). We use $f_A$ instead of $f$ because we will have to scale the results in the end to ensure their sum equals $1$. All formulas so far also hold true for $f_A$ and we have $f = A \cdot f_A$ for some constant $A \in \mathbb{R}$ to be found. (In the multidimensional case, more initial values need to be used and thus more constants need to be considered.)

Henceforward, the entire distribution over the $y$-axis can be evaluated using \eqref{eq::recursion}. Evaluation must stop at some $\hat{y}$ due to practical limitations.
Using the propagation along the diagonals via \eqref{eq::diagonal_attenuation}, the entire $(y>x)$-half of the state-space can be evaluated as well. Once this is done, the boundary distribution on the $x$-axis, $f_A(x,1)$, follows as
\begin{equation}
    \label{eq::get_x_last}
    f_A(x+1,1) = q \sum_{y=y'(x)}^{\hat{y}} f_A(x,y)
\end{equation}
Evaluation of the boundary distribution $f_A(x,1)$ has to end at $\hat{x} = \left \lfloor \frac{q \hat{y}}{p} \right \rfloor+1 \leq \hat{y}$. This is the $x$-component of the last state that is reachable from any prior evaluated states. Subsequently, the probability of the missing diagonals can be determined, after which $f$ follows from $f_A$ through normalization to $1$ (i.e. finding $A$). 
Fig. \ref{fig::mw_triple} shows the resulting distribution for the parameters $p = 0.6$ and $q=0.2$.

Remark: For $\frac{p}{q} =: \kappa \in \mathbb{N}$ and $y = \kappa m +1$ and $m = 1,2,\dots$ (i.e. looking only at every $\kappa$-th value of $y$), \eqref{eq::recursion} simplifies to
\begin{equation}
    \label{eq::panto}
    f(1,\kappa m+1)
    =
    \bar{p}
    f(1,\kappa m)
    +
    p \bar{q}^{m-1}f(1,(\kappa-1) m + 1) 
\end{equation}
Every $y$ value, not covered by this recursion can be expressed by exponential attenuation. 
A solution for this difference equation would allow for an analytical description of the distribution over the entire state-space. However, such a difference equation with \textit{proportional delay} is notoriously hard to solve, especially since \eqref{eq::panto} exhibits a non-constant factor in front of the delayed term. The most recent results on similar equations can be found in \cite{Kundrat2006} and references therein. In further abstraction, \eqref{eq::panto} is the discrete version of the so called Pantograph equation, first published in \cite{Ockendon1971}. However, due to the discrete argument, the difference equation does only hold for certain arguments (those which abide to the corresponding divisibility) whereas the Pantograph equation holds for all its real arguments. Therefore, most results on the Pantograph equation are not transferable to our problem.

\subsection{Evaluation of the State-Space Distribution under the Optimal Policy for $n=2$}

\begin{figure}
    \centering
    \footnotesize
    \includegraphics[width = 0.75\columnwidth]{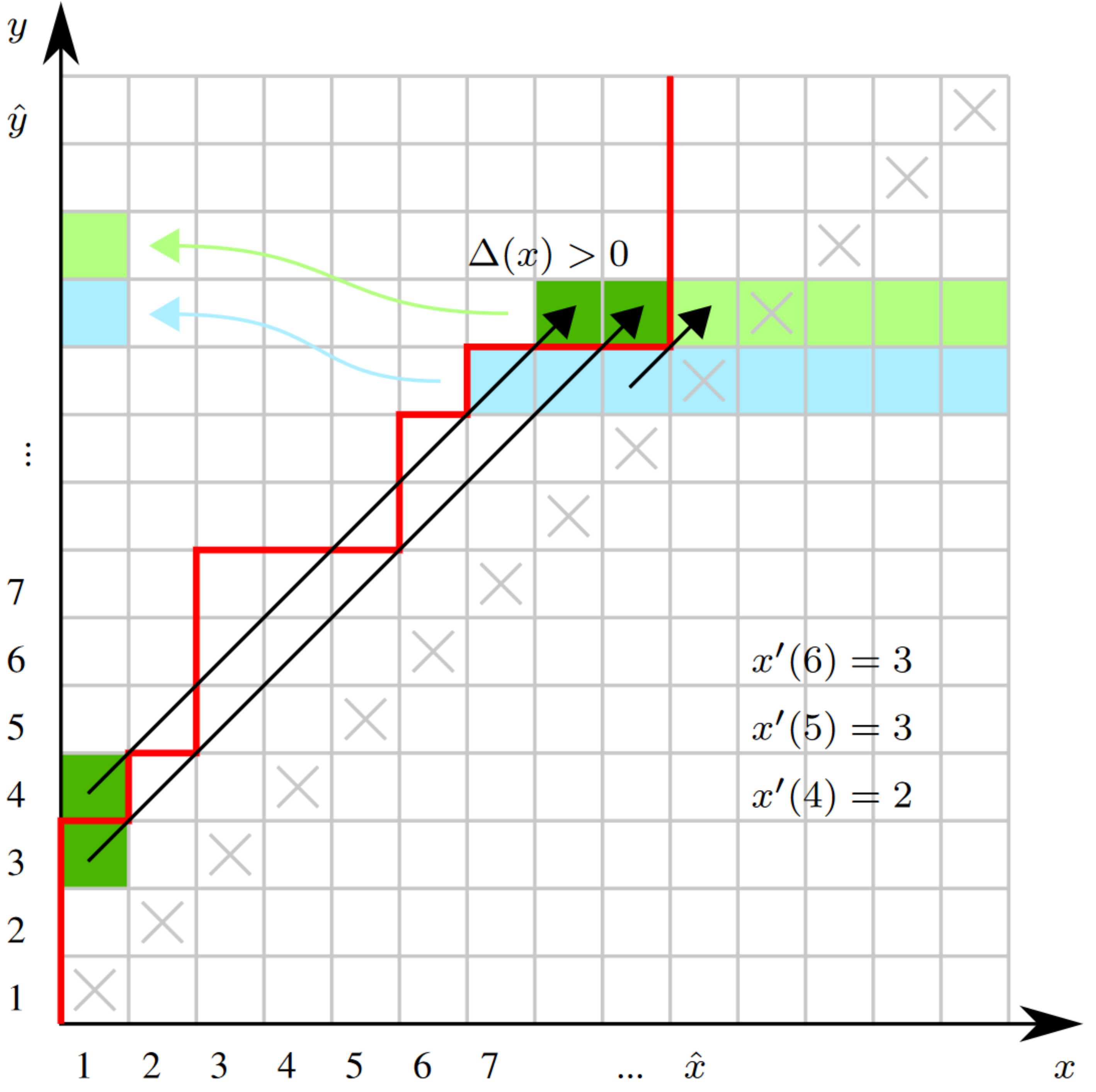}
    \caption{General root-state recursion in 2-dim. state-space. The dark-green states are dropped compared to the previous line (blue). Their contribution has to be subtracted which requires involvement of their root states (also dark-green).}
    \label{fig::gen_00}
\end{figure}

Still staying in 2 dimensions, we will now discuss a more general example in which the policy is defined only numerically, i.e. $d(a)$ is given by a matrix. Once again we will identify $a^{1}$ and $a^{2}$ with $x$ and $y$, and $p^{1}$ and $p^{2}$ with $p$ and $q$, respectively. The policy shall now be defined by \figref{fig::gen_00}: For all states above the separating line (red line), the policy tries to reset the $y$ component, whereas for all states beneath, it tries to reset the $x$ component.

Since the function $x'(y)$ is now not given by a mathematical expression, we need to consider the change in $x'(y)$ for each step separately. Given the pure exponential attenuation of the root-states' probability in the case of $x'(y) = x'(y-1) + 1$, we define
\begin{equation}
    \label{eq::gen_con}
    \Delta(y) = x'(y) - x'(y-1) - 1
\end{equation}
With that, the following more general recursion holds
\begin{equation}
    \label{eq::gen_rec}
\begin{gathered}
    f(y+1,1) = \bar{p} f(y,1) + p \cdot \dots
    \\[2ex]
    \dots
    \begin{cases}
        \displaystyle
        \sum_{i=0}^{-\Delta(y)-1} f(y,x'(y)+i) &  \text{if } \Delta(y) < 0
        \\[4ex]
        \displaystyle
        0 &  \text{if } \Delta(y) = 0
        \\[2ex]
        \displaystyle
        -\sum_{i=1}^{\Delta(y)} f(y,x'(y)-i) &  \text{if } \Delta(y) > 0
    \end{cases}
\end{gathered}
\end{equation}
where root-states can be substituted via
\begin{equation}
    f(y,x'(y) \pm i) = f(y-x'(y) \mp i +1 ,1) D(y,x'(y) \pm i)
\end{equation}
With an initial parameter, the state-space can thus be evaluated using Algorithm \ref{alg::algo}.

\begin{algorithm}
    \caption{Pseudo-Algorithm for Two-Agent System}
    \label{alg::algo}
    \begin{algorithmic}
        \Require $f$ (2D-Matrix)
        \Require $\hat{y}$ (Size of $y$-dimension of matrix $f$)
        \State $f(1,2) \gets 1$
        \State $y \gets 2$
        \While{$y < \hat{y}-1$}
            \State Evaluate $\Delta(y)$, using \eqref{eq::gen_con}
            \State Evaluate $f(1,y+1)$ using \eqref{eq::gen_rec}
            \State Evaluate states, diagonal to $f(1,y+1)$ using \eqref{eq::diagonal_attenuation}
            \State $y \gets y+1$
        \EndWhile
        \State$x \gets 2$
        \While{$x \leq \lfloor \frac{q}{p}y \rfloor + 1 $}
            \State Evaluate $f(x,1)$, using \eqref{eq::get_x_last}
            \State $x \gets x+1$
        \EndWhile
        \State Normalize matrix $f$
    \end{algorithmic}
\end{algorithm}

\begin{figure}
    \centering
    \footnotesize
    \includegraphics[width = 0.75\columnwidth]{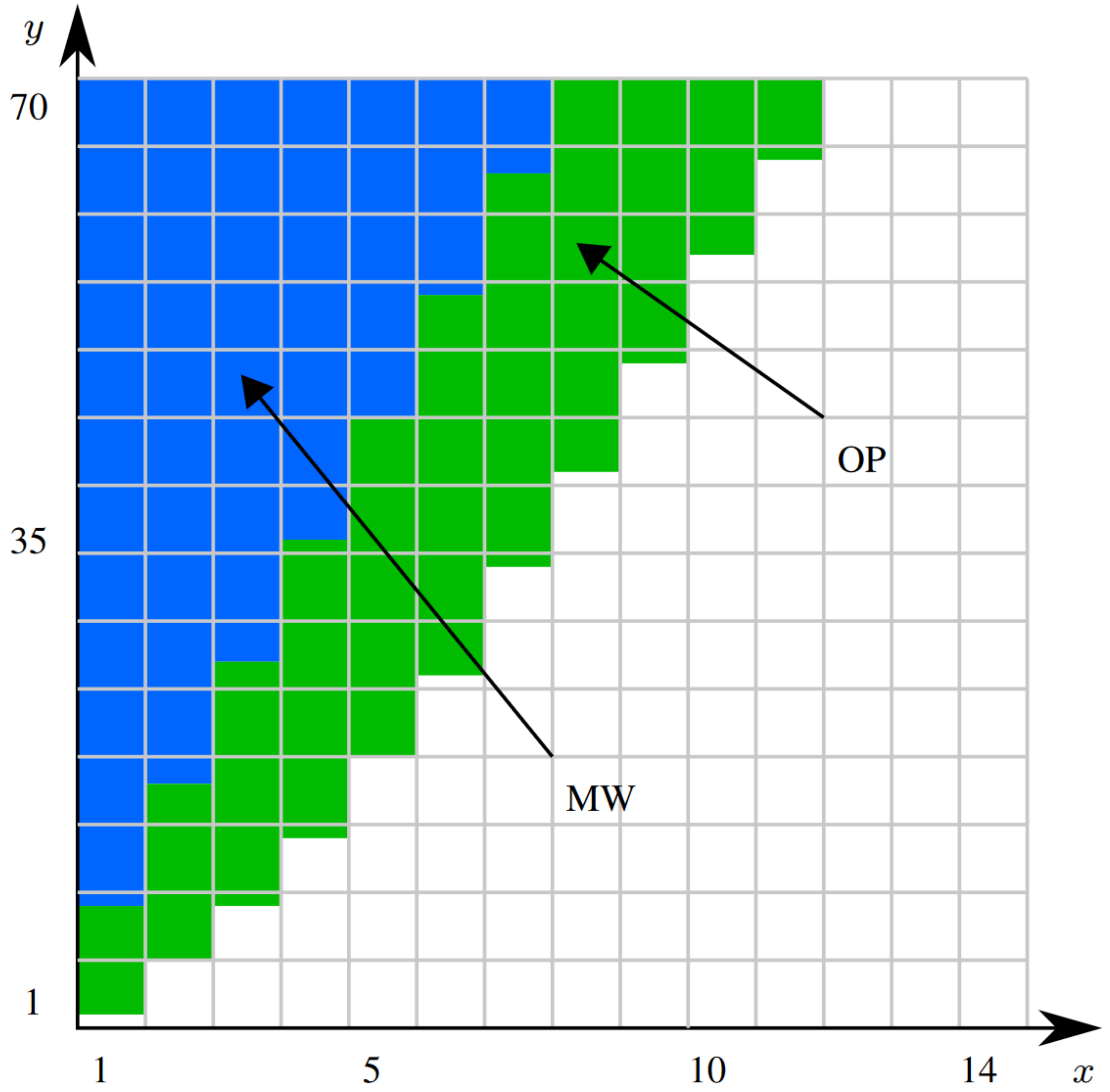}
    \caption{Truncated decision matrix for MW (blue) and OP (blue + green). If the current state resides in the colored region, each policy tries to reset the $y$-component of the AoI with success probability $q=0.1$, otherwise the $x$-component with probability $p=0.9$. Compared with MW, OP tries to reset the $y$-component (corresponding to the weak communication link) much more often.}
    \label{fig::dec_maps}
\end{figure}

We can employ the presented method on the \textit{optimal} control policy (OP). Though to the best of our knowledge, an analytical expression is not yet found for this optimal policy, it is easily obtained by policy-iteration methods in form of a decision matrix like the one in \figref{fig::gen_00} (the separating line corresponds to the policy). For the case of $p=0.9$ and $q=0.1$, the decision matrices of the MW policy and the OP policy are depicted in \figref{fig::dec_maps}. The change in the resulting state-space distributions is visualized in \figref{fig::MW_vs_OP}.

Doing this evaluation for every combination of $p=\{0.1 ,\dots 0.9\}$ and $q= \{0.1 \dots 0.9\}$, allows one to yield the comparison in performance of MW and OP policy. Therefor, every state probability is multiplied with its corresponding AoI value (identical to its $L^0$-norm). The sum over all states then gives the average AoI under the policy. As can be seen in \figref{fig::performance}, OP has a performance gain (less average AoI) of about 15\% over MW in the case of $p=0.9$ and $q=0.1$. The closer the probabilities are, the smaller this advantage gets, since the policies coincide for $p=q$.

\usetikzlibrary{backgrounds}
\usepgfplotslibrary{colormaps}

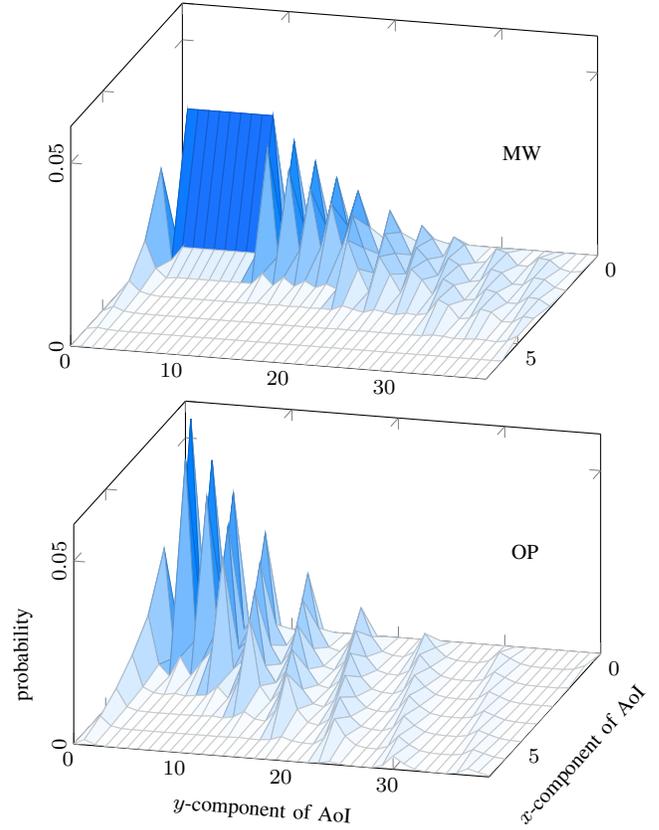
\begin{figure}
    \centering
    \footnotesize
    \begin{tikzpicture}[x = 1mm, y = 1mm]
        \begin{axis}[
            scale only axis,
            axis background/.style={fill=white},
            colormap/cool,
            view = {105}{30},
            width = 70mm,
            height = 50mm,
            font = \footnotesize,
            ztick={0,0.05},
            zmin={0}, zmax = {0.06},
            zticklabels={ $0$ , $0.05$ },
            z tick label style={rotate=90,anchor=south},
            ztick scale label code/.code={},
            ]
        \addplot3 [surf, mesh/ordering=y varies] table {data_proma_p09_q01_mw.txt};
        \end{axis}%
        \node at (60,30) {MW};
    \end{tikzpicture}
    \begin{tikzpicture}[x = 1mm, y = 1mm]
        \begin{axis}[
            scale only axis,
            colormap/cool,
            view = {105}{30},
            width = 70mm,
            height = 50mm,
            font = \footnotesize,
            xlabel style = {sloped},
            xlabel={$x$-component of AoI},
            ztick={0,0.05},
            zmin={0}, zmax = {0.06},
            zticklabels={ $0$ , $0.05$ },
            ztick scale label code/.code={},
            ylabel style = {sloped},
            ylabel={$y$-component of AoI},
            zlabel={probability},
            z tick label style={rotate=90,anchor=south},
            ]
        \addplot3 [surf, mesh/ordering=y varies] table {data_proma_p09_q01_op.txt};
        \end{axis}%
        \node at (60,30) {OP};
    \end{tikzpicture}
    \caption{State-space probability distribution for transmission success probabilities $p=0.9$ and $q=0.1$ under MW (top) and OP policy (bottom). The wave-like distribution under MW can be attributed to equation \eqref{eq::panto}. An intuitive explanation is derived in \figref{fig::wave}.}
    \label{fig::MW_vs_OP}
\end{figure}

\begin{figure}
    \centering
    \footnotesize
    \includegraphics[width = 0.75\columnwidth]{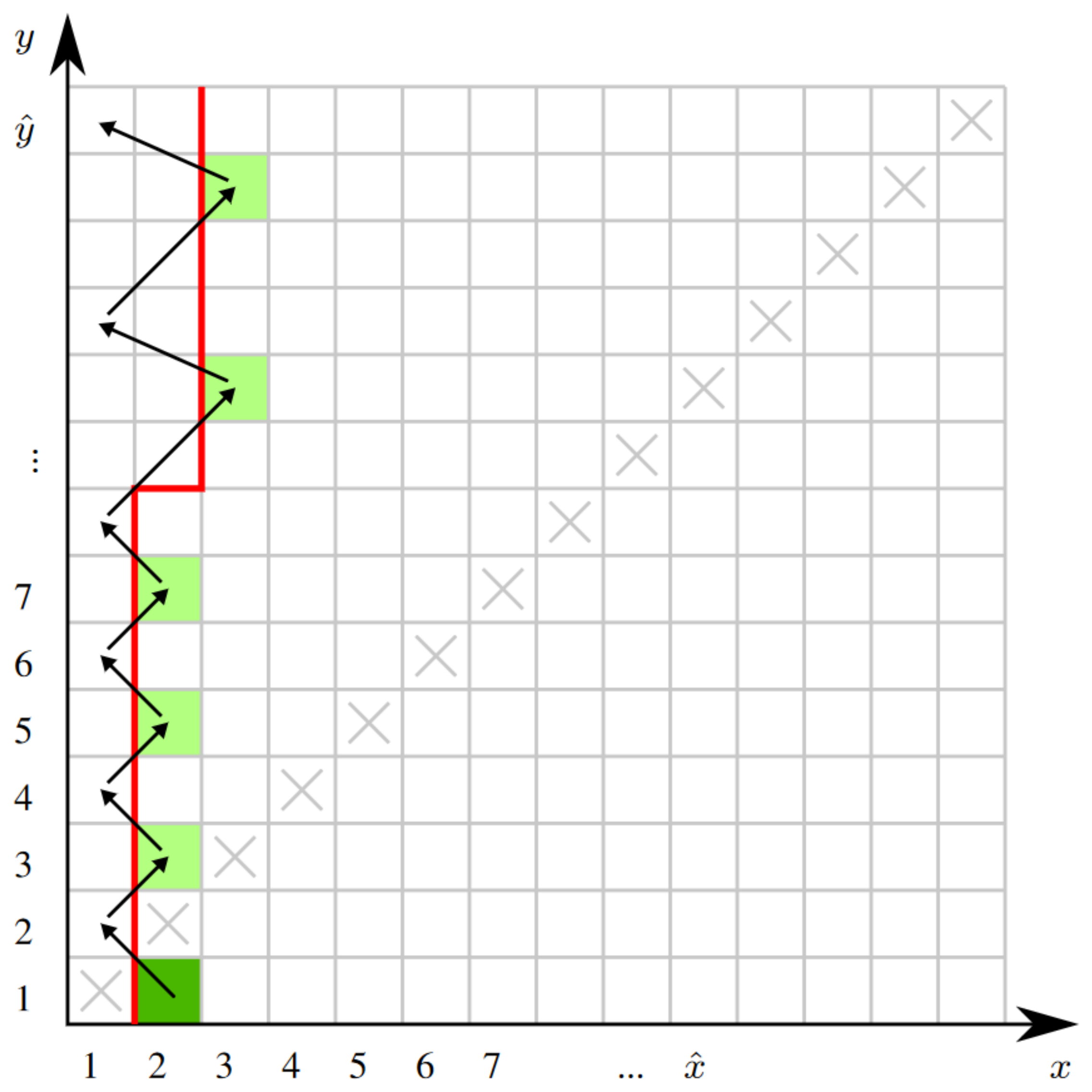}
    \caption{Intuitive Explanation for wave-like form of the probability distribution in \figref{fig::MW_vs_OP}: Let the probability of a successful reset of $x$ and $y$ be high and low, respectively. Then, if the dark-green state is highly probable, so are the light-green states. And even if a $y$ reset is to succeed during the indicated evolution, there is a high probability of reaching the dark-green state again.}
    \label{fig::wave}
\end{figure}

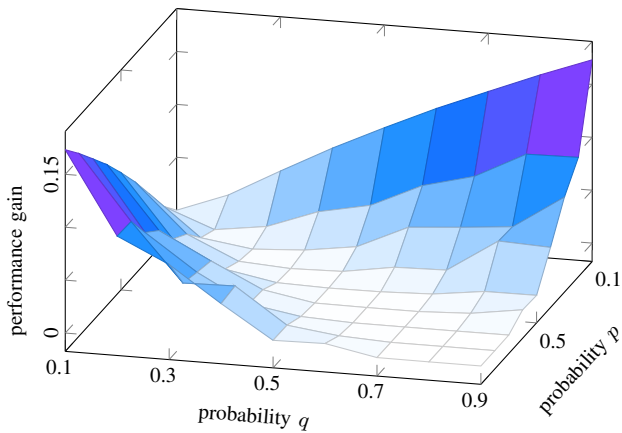
\begin{figure}
    \centering
    \scriptsize
    \begin{tikzpicture}
    \begin{axis}[
        scale only axis,
        colormap/cool,
        view = {105}{30},
        width = 70mm,
        height = 50mm,
        font = \footnotesize,
        xlabel style = {sloped},
        xtick = {0,4},
        xticklabels = {0.1,0.5},
        ytick = {0,2,4,6,8},
        yticklabels = {0.1,0.3,0.5,0.7,0.9},
        ztick = {0 , 0.05 , 0.1 , 0.15},
        zticklabels = {0 ,  ,  , 0.15},
        xlabel={probability $p$},
        ylabel style = {sloped},
        ylabel={probability $q$},
        zlabel={performance gain},
        z tick label style={rotate=90,anchor=south},
        ]
    \addplot3 [surf, mesh/ordering=y varies] table {data_mw_vs_op.txt};
    \end{axis}%
    \end{tikzpicture}
    \caption{Increased performance of OP over MW in \% of average AoI drop. Probabilities $p$ and $q$ represent transmission success probabilities between the agents. For $p=q$, MW coincides with OP. For differing values of $p$ and $q$, OP performs better.}
    \label{fig::performance}
\end{figure}

\section{Noteworthy Observations}
\label{sec::not_obs}

Looking at the results, there is a surprising observation visualized in \figref{fig::op_curve}. As evaluated by the proposed algorithm and cross checked with Monte-Carlo-Simulations, there does not seem to be a smooth decay of the average AoI when increasing the success probability $p$ under the MW policy. Setting $q = 0.1$ and looking at the values for $p=0.94$ and $p=0.95$, the average AoI increases while $p$ increases. I.e. the system performance decreases though the network reliability strictly increases. Using only Monte-Carlo-Simulations to determine the average AoI, one would probably misinterpret such a finding as sample noise. Our algorithm, however, proves that this seems to be a genuine effect.

An explanation for this effect might be found in \figref{fig::dec_maps}. The figure shows that MW gives much less priority to the activation of the weak link ($y$-component of the AoI), compared to OP. Increasing $p$ from $p=0.94$ to $p=0.95$ will lessen MW's priority on activating the weak link even more (since the weight of the strong link becomes even greater). Given that the state-space is discrete, the slight change in $p$ will make some of MW's state-dependent decisions flip from "activate weak link" to "activate strong link" (equal to slightly less blue area in \figref{fig::dec_maps}). These non-smooth flips, together with the fact that for optimal behavior, a policy should rather try to activate the weak link more often might cause the observed behavior.

\pgfplotstableread{data_non_smooth_decay.txt}{\nonsmoothtable}

\pgfplotsset{compat=1.17}

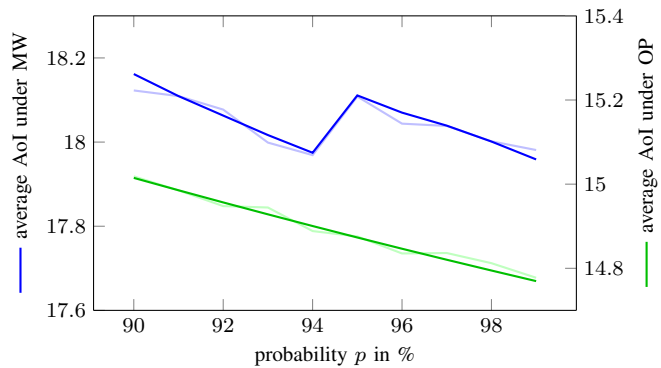
\begin{figure}
    \centering
    \footnotesize
    \begin{tikzpicture}
        \begin{axis}[
            width = 80mm,
            height = 55mm,
            xlabel = {probability $p$ in \%},
            ylabel = {\ref{plot::mw} average AoI under MW},
            axis y line*=left,
            ymin = {17.6},
            ymax = {18.3}
            ]
            \addplot[ blue!25!white, thick ] table [x = {x} , y = {mw_incma}] {\nonsmoothtable};
            \addplot[ blue, thick ] table [x = {x} , y = {mw_proma}] {\nonsmoothtable};
            \label{plot::mw}
        \end{axis}%
        \begin{axis}[
            width = 80mm,
            height = 55mm,
            ylabel = {\ref{plot::op} average AoI under OP},
            axis y line*=right,
            axis x line=none,
            ymin = {14.7},
            ymax = {15.4}
            ]
            \addplot[ green!25!white, thick ] table [x = {x} , y = {op_incma}] {\nonsmoothtable};
            \addplot[ green!75!black, thick ] table [x = {x} , y = {op_proma}] {\nonsmoothtable};
            \label{plot::op}
       \end{axis}%
    \end{tikzpicture}
    \caption{Average AoI for $q = 0.1$ and $p = 0.9 ,\dots 0.99$. Exact values using the algorithm (dark-blue / dark-green) and values obtained with Monte-Carlo-Simulation ($10^7$ steps) (light-blue and light-green). The non-smooth decay under MW stands out.}
    \label{fig::op_curve}
\end{figure}

\section{Conclusion}

We presented an algorithm with which, for the first time, it is possible to yield an exact numerical comparison between the AoI-optimizing policy and the MaxWeight policy in terms of average AoI in a two-agent network. For this case, we show the connection to a difference equation with proportional delay (related to the infamous Pantograph equation) and present simulation results for different transmission parameters. Our algorithm is based on the evaluation of the probability distribution over the state-space, facilitated by the special properties of the AoI process. It is readily applicable to any causal policy and can also be extended to the case of arbitrary many agents. In contrast to Monte-Carlo-Simulations, our algorithm allows for a much faster evaluation and additionally is much more exact, facilitating the evaluation of higher order moments of the underlying stochastic AoI process.

\bibliographystyle{IEEEtran}
\bibliography{library}

\begin{thebibliography}{10}
\providecommand{\url}[1]{#1}
\csname url@samestyle\endcsname
\providecommand{\newblock}{\relax}
\providecommand{\bibinfo}[2]{#2}
\providecommand{\BIBentrySTDinterwordspacing}{\spaceskip=0pt\relax}
\providecommand{\BIBentryALTinterwordstretchfactor}{4}
\providecommand{\BIBentryALTinterwordspacing}{\spaceskip=\fontdimen2\font plus
\BIBentryALTinterwordstretchfactor\fontdimen3\font minus
  \fontdimen4\font\relax}
\providecommand{\BIBforeignlanguage}[2]{{%
\expandafter\ifx\csname l@#1\endcsname\relax
\typeout{** WARNING: IEEEtran.bst: No hyphenation pattern has been}%
\typeout{** loaded for the language `#1'. Using the pattern for}%
\typeout{** the default language instead.}%
\else
\language=\csname l@#1\endcsname
\fi
#2}}
\providecommand{\BIBdecl}{\relax}
\BIBdecl

\bibitem{Kadota2016}
I.~Kadota, E.~Uysal-Biyikoglu, R.~Singh, and E.~Modiano, ``{Minimizing the Age
  of Information in Broadcast Wireless Networks},'' \emph{Fifty-fourth Annual
  Allerton Conference}, 2016.

\bibitem{Hsu2017}
Y.~P. Hsu, E.~Modiano, and L.~Duan, ``{Age of information: Design and analysis
  of optimal scheduling algorithms},'' \emph{IEEE International Symposium on
  Information Theory - Proceedings}, 2017.

\bibitem{Hsu2020}
------, ``{Scheduling Algorithms for Minimizing Age of Information in Wireless
  Broadcast Networks with Random Arrivals},'' \emph{IEEE Transactions on Mobile
  Computing}, 2020.

\bibitem{Kadota2019}
I.~Kadota, A.~Sinha, and E.~Modiano, ``{Scheduling algorithms for optimizing
  age of information in wireless networks with throughput constraints},''
  \emph{IEEE/ACM Transactions on Networking}, 2019.

\bibitem{Kadota2018}
I.~Kadota, A.~Sinha, E.~Uysal-Biyikoglu, R.~Singh, and E.~Modiano,
  ``{Scheduling policies for minimizing age of information in broadcast
  wireless networks},'' \emph{IEEE/ACM Transactions on Networking}, 2018.

\bibitem{Kadota2018a}
I.~Kadota, A.~Sinha, and E.~Modiano, ``{Optimizing Age of Information in
  Wireless Networks with Throughput Constraints},'' \emph{Proceedings - IEEE
  INFOCOM}, 2018.

\bibitem{Arafa2017}
A.~Arafa and S.~Ulukus, ``{Age-minimal transmission in energy harvesting
  two-hop networks},'' \emph{GLOBECOM IEEE Global Communications Conference -
  Proceedings}, 2017.

\bibitem{Yates2018}
R.~D. Yates, ``{Age of information in a network of preemptive servers},''
  \emph{INFOCOM 2018 - IEEE Conference on Computer Communications Workshops},
  2018.

\bibitem{Farazi2019}
S.~Farazi, A.~G. Klein, and D.~{Richard Brown}, ``{Fundamental bounds on the
  age of information in multi-hop global status update networks},''
  \emph{Journal of Communications and Networks}, 2019.

\bibitem{Hahn2021}
J.~Hahn, R.~Schoeffauer, G.~Wunder, and O.~Stursberg, ``{Using AoI Forecasts in
  Communicating and Robust Distributed Model-Predictive Control},'' \emph{IEEE
  Transactions on Control of Network Systems}, 2021.

\bibitem{Ayan2019}
O.~Ayan, M.~Vilgelm, M.~Kl{\"{u}}gel, S.~Hirche, and W.~Kellerer,
  ``{Age-of-information vs. Value-of-information scheduling for cellular
  networked control systems},'' \emph{arXiv}, 2019.

\bibitem{Sinha2019}
D.~Sinha and R.~Roy, ``{Scheduling Status Update for Optimizing Age of
  Information in the Context of Industrial Cyber-Physical System},'' \emph{IEEE
  Access}, 2019.

\bibitem{Kundrat2006}
P.~Kundr{\'{a}}t, ``{On the Asymptotics of the Difference Equation with a
  Proportional Delay},'' \emph{Opuscula Mathematica}, 2006.

\bibitem{Ockendon1971}
J.~Ockendon and A.~Tayler, ``{The dynamics of a current collection system for
  an electric locomotive},'' \emph{Proceedings of the Royal Society of London.
  A. Mathematical and Physical Sciences}, 1971.

\end{thebibliography}

\end{document}